\documentclass[a4paper, 12pt]{article}
\usepackage{amsmath, amssymb, graphics}
\newcommand{\mathsym}[1]{{}}
\newcommand{\unicode}[1]{{}}

\catcode`\@=11

\@addtoreset{equation}{section} 

\def\dddot#1{\mathinner{\buildrel\vbox{\kern5pt\hbox{...}}\over{#1}}}
\def\ddddot#1{\mathinner{\buildrel\vbox{\kern5pt\hbox{....}}\over{#1}}}

\begin{document}

\begin {center}
{\Large Lie symmetries and similarity solutions for the generalized Zakharov equations}\\[3 mm]
{\small K. Krishnakumar$^{1}$, A. Durga Devi$^{2}$ and A. Paliathanasis}$%
^{3,4}${\small \\[3mm]
$^{1}$ Department of Mathematics, Srinivasa Ramanujan Centre, SASTRA Deemed
to be University, Kumbakonam 612 001, India.\\[0pt]
$^{2}$ Department of Physics, Srinivasa Ramanujan Centre, SASTRA Deemed to be University, Kumbakonam 612 001, India.}\\[0pt]
$^{3}${\small Institute of Systems Science, Durban University of Technology,
Durban 4000, South Africa}\\[0pt]
$^{4}${\small Instituto de Ciencias F\'{\i}sicas y Matem\'{a}ticas,
Universidad Austral de Chile, Valdivia 5090000, Chile}
\end{center}

\begin{center}
\textbf{ABSTRACT}

The theory of Lie point symmetries is applied to study the generalized Zakharov system with two unknown parameters. The system reduces into a
three-dimensional real value functions system, where we find that admits
five Lie point symmetries. From the resulting point, we focus on these which provide travel-wave similarity transformation. The reduced system can be integrated while we remain with a system of two second-order nonlinear ordinary differential equations. The parameters of the latter system are classified in order the equations to admit Lie point symmetries. Exact travel-wave solutions are found, while the generalized Zakharov system can be described by the one-dimensional Ermakov-Pinney equation.
\end{center}



\strut\hfill

\section{Introduction}

One side, it is usual practice for many years to develop a mathematical
model to study the nature of the real life phenomena. In which differential
equations are playing a crucial role to describe the phenomena very clearly.
Another side, many researchers are trying to solve the differential
equations by using various methods. Especially, Sophus Lie, during the
period $1872-1899$ \cite{Lie 70 a, Lie 70 b, Lie 70 c, Lie 67 a, Lie 71 a,
Lie 77 a, Lie 12 a}, developed and applied Symmetry Analysis method
successfully to solve the differential equations. Without any ansatz, the
Symmetry Analysis directly can apply in a systematic way to derive solutions
of differential equations. Later, this theory was exploited by the Russian
school with L. V. Ovsiannikov \cite{Ovsyannikov 58 a, Ovsyannikov 59 a,
Ovsiannikov 82 a} that since 1960 over the explicit construction of
solutions of any sort of problems, even complicated, of mathematical
physics. During the last few decades, Lie$^{\prime }$s theory has been given
much attention on both theoretical and applied point of view \cite{Bluman 69
a, Bluman 69 b, BluC74, Bluman 88 a, Bluman 89 a, Ibragimov 83 a, Ibragimov
99 a, nailio, Ibraghimov, Olver 77 a, Olver 86 a, Olver 93 a, Olver 02
a,Olver 03 a, Andriopoulos 06 a, Leach 78 a, Leach 80 e, Leach 80 f, Leach
81 a, Leach 03 c}.\hfill

Algorithm of Lie's theory to differential equations is completely
algorithmic involves tedious calculations even for linear differential
equations with constant coefficients. To overcome the difficulties we have
access to powerful Computer Algebra Systems (CAS) like Maple and Mathematica
(commercial), etc. which are enabling us to do the calculations rapidly%
\footnote{%
In this work, for the calculation of the symmetries we use the Mathematica
add-on Sym \cite{Dimas 05 a, Dimas 06 a, Dimas 08 a, Andriopoulos 09 a}.}.
In the
 past decades, many types of symmetries have been proposed in the literature such as approximate symmetries \cite{Baikov 89, Ibraghimov 09}%
, generalized symmetries \cite{Olver 93 a, Ovsiannikov 82 a}, and nonlocal
symmetries \cite{Olver 93 a, Bluman 89 a, Olver 03 a, Leach 07} to quote a
few.

The novelty of the Lie's theory is that provides a systematic way to treat
nonlinear differential equations. The main application of Lie's theory are
summarized to: the determination of similarity transformations which are
used to reduce the differential equation; to determine invariant solutions,
also known as similarity solutions, construct conservation laws and write
linearize differential equations \cite{Olver 93 a, Bluman 89 a, Olver 03 a,
Leach 07}.

Applications of Lie symmetries cover a wide range of physical and natural
systems, from classical mechanics \cite{app1,app2}, fluid dynamics \cite%
{app3,app4}, plasma physics \cite{app4a}, optics \cite{app5}, gravitational
physics \cite{app6,app7}, financial mathematics \cite{kk1, kk2} and other areas of applied mathematics.

In this work are interesting on the symmetry analysis for the generalized
Zakharov equations \cite{zakh0} of plasma physics \cite{zakh01}
\begin{eqnarray}
iU_{t}+U_{xx}-UV-(|U|^{2n})U &=&0  \label{ss.01} \\
V_{tt}-V_{xx}-({|U|^{2m}})_{xx} &=&0  \label{ss.02}
\end{eqnarray}%
where the complex function $U\left( t,x\right) ~$is the envelope of the
high-frequency electric field and the real function $V\left( t,x\right) $%
~plasma density measured from its equilibrium value. Parameters $n,$ $m$ are
arbitrary in our consideration, while they describe the nonlinear
self-interaction in the high-frequency subsystem which corresponds to a
self-focusing effect in plasma physics \cite{zakh1}. For the system (\ref%
{ss.01}), (\ref{ss.02}) travel-wave solutions were studied in \cite{zakh1}.

In the following we focus on the Lie point symmetries for the system (\ref%
{ss.01}), (\ref{ss.02}) while the free parameters $n,m$ will be determined
by Lie's theory, such that the resulting system to admit additional
symmetries, that approach is inspired by the Ovsiannikov's classification
scheme. The plan of the paper is as follows.

In Section 2 we briefly discuss the theory of Lie point symmetries. The Lie
point symmetries of the generalized Zakharov system are derived in Section
3. We find that the system admits a five dimensional Lie algebra, we use the
Lie symmetries to study the existence of travel wave solutions. The reduced
system is classified according to the admitted Lie point symmetries.
Surprisingly, we find the generalized Zakharov equations can be reduced into
the Ermakov-Pinney equation. Finally, in Section 4 we summarize our results
and we draw our conclusions.

\section{Lie's Theory}

Suppose consider an equation
\begin{equation}
\phi
(t,x,y,u,v,u_{t},v_{t},u_{x},v_{x},u_{y},v_{y},u_{tt},v_{tt},u_{tx},v_{tx},u_{ty},v_{ty},u_{xx},u_{xy},...)=0,
\label{1.1}
\end{equation}%
where $t,x,y$ are the set of independent variables and $u,v$ are dependent
variables. Infinitesimal point transformation for each variables is defined
as in the following manner,
\begin{equation*}
\tilde{t}(t,x,\ y,\ \epsilon )=t+\epsilon \xi ^{1}(t,x,\ y)+\circ (\epsilon
^{2})=t+\epsilon Xt+\circ (\epsilon ^{2}),
\end{equation*}%
\begin{equation*}
\tilde{x}(t,x,\ y,\ \epsilon )=x+\epsilon \xi ^{2}(t,x,\ y)+\circ (\epsilon
^{2})=x+\epsilon Xx+\circ (\epsilon ^{2}),
\end{equation*}%
\begin{equation*}
\tilde{y}(t,x,\ y,\ \epsilon )=y+\epsilon \xi ^{3}(t,x,\ y)+\circ (\epsilon
^{2})=y+\epsilon Xy+\circ (\epsilon ^{2}),
\end{equation*}%
\begin{equation*}
\tilde{u}(t,x,\ y,\ \epsilon )=u+\epsilon \eta ^{1}(t,x,\ y)+\circ (\epsilon
^{2})=u+\epsilon Xu+\circ (\epsilon ^{2}),
\end{equation*}%
\begin{equation*}
\tilde{v}(t,x,\ y,\ \epsilon )=v+\epsilon \eta ^{2}(t,x,\ y)+\circ (\epsilon
^{2})=v+\epsilon Xv+\circ (\epsilon ^{2}),
\end{equation*}%
where $X$ is called {\emph{infinitesimal generator}} which is denoted by%
\newline
\begin{equation*}
X=\xi ^{1}(t,x,\ y)\partial _{t}+\xi ^{2}(t,x,\ y)\partial _{x}+\xi
^{3}(t,x,\ y)\partial _{y}+\eta ^{1}(t,x,\ y)\partial _{u}+\eta ^{2}(t,x,\
y)\partial _{v}
\end{equation*}%
Based of the theory the invariant condition for (\ref{1.1}) is given by
\begin{equation*}
\phi (t,x,y,u,v)=\phi (\tilde{t},\tilde{x},\tilde{y},\tilde{u},\tilde{v})
\end{equation*}%
It is well known that one can reduce the order of the differential equations
as well as number of independent variables by using the infinitesimal
generator which is known as symmetries of (\ref{1.1}).

Therefore, if $X$ is a Lie point symmetry for equation $\phi \equiv 0$, then
the following condition is true%
\begin{equation*}
X^{\left[ k\right] }\phi =\lambda \phi ~,~
\text{mod}\phi =0
\end{equation*}%
where $\lambda $ is an arbitrary function, and $X^{\left[ k\right] }$ is the
k-th extension of $X$ in the jet-space.

\section{Symmetries of Generalized Zakharov Equation}

We write the dynamical system (\ref{ss.01}), (\ref{ss.02}) as follows
\begin{eqnarray}
-q_{t}+p_{xx}-pv-p(p^{2}+q^{2})^{n} &=&0  \notag  \label{2.2} \\
p_{t}+q_{xx}-qv-q(p^{2}+q^{2})^{n} &=&0  \notag \\
v_{tt}-v_{xx}-2m(p^{2}+q^{2})^{m-1}(pp_{xx}+qq_{xx}+{p_{x}}^{2}+{q_{x}}%
^{2})- &&  \notag \\
4m(m-1)(p^{2}+q^{2})^{m-2}(pp_{x}+qq_{x})^{2} &=&0
\end{eqnarray}%
after substituting the transformations $U=p+iq$ and $V=v.$

The latter real valued system admits five Lie point symmetries

\begin{equation}
\begin{array}{lll}
\Gamma _{1} & = & \partial _{t}, \\
\Gamma _{2} & = & \partial _{x}, \\
\Gamma _{3} & = & p\partial _{q}-q\partial _{p}, \\
\Gamma _{4} & = & qt\partial _{p}-pt\partial _{q}+\partial _{v}, \\
\Gamma _{5} & = & -qt^{2}\partial _{p}+pt^{2}\partial _{q}-2t\partial _{v}.%
\end{array}%
\end{equation}

On the other hand, if we use the variables $U=Re^{i\theta }$, the Lie
symmetries are simplified as
\begin{equation}
\begin{array}{lll}
\Gamma _{1} & = & \partial _{t}, \\
\Gamma _{2} & = & \partial _{x}, \\
\Gamma _{3} & = & \partial _{\theta }, \\
\Gamma _{4} & = & t\partial _{\theta }+\partial _{v}, \\
\Gamma _{5} & = & t^{2}\partial _{\theta }-2t\partial _{v}.%
\end{array}%
\end{equation}

In the following we continue our analysis by applying the Lie point
symmetries which provide travel-wave solutions.

\subsection{Travel Wave solution}

The traveling wave solution of (\ref{2.2}) can be constructed by taking
linear combination $\Gamma _{1}$ and $\Gamma _{2}.$ The new canonical
variable is $r=x-ct.$ Therefore, the system of PDE (\ref{2.2}) reduces to
system of ODE as follows

\begin{eqnarray}  \label{2.3}
p^{\prime \prime }+cq^{\prime }-p[v+(p^{2}+q^{2})^{n}] &=&0  \label{2.7} \\
q^{\prime \prime }-cp^{\prime }-q[v+(p^{2}+q^{2})^{n}] &=&0  \label{2.8} \\
(c^{2}-1)v^{\prime \prime }-2m(p^{2}+q^{2})^{m-1}(pp^{\prime \prime
}+qq^{\prime \prime }+{p^{\prime }}^{2}+{q^{\prime }}^{2})- &&  \notag \\
4m(m-1)(p^{2}+q^{2})^{m-2}(pp^{\prime }+qq^{\prime })^{2} &=&0  \label{2.4}
\end{eqnarray}%
where prime represents the derivative with respect to $r.$

The solution of (\ref{2.4}) is given by

\begin{eqnarray}  \label{2.6}
v&=&\frac{(p^2+q^2)^m}{c^2-1}+(c_1r+c_2)
\end{eqnarray}

By substituting the expression of $v$ in equations (\ref{2.7}) and (\ref{2.8}%
) we have

\begin{eqnarray}
p^{\prime\prime}+c q^\prime&=&p\left(\frac{(p^2+q^2)^m}{c^2-1}
+(c_1r+c_2)+(p^2+q^2)^n\right)  \label{2.9} \\
q^{\prime\prime}-c p^\prime&=&q\left(\frac{(p^2+q^2)^m}{c^2-1}
+(c_1r+c_2)+(p^2+q^2)^n\right)  \label{2.10}
\end{eqnarray}

The above equations in general are having a rotational symmetry which is
symmetry $\Gamma _{3}$. \ Now we have to solve only the equations (\ref{2.9}%
) and (\ref{2.10}).

Divide equations (\ref{2.9}) and (\ref{2.10}) then we have

\begin{equation}
\frac{p^{\prime \prime }+cq^{\prime }}{q^{\prime \prime }-cp^{\prime }}=%
\frac{p}{q}
\end{equation}

This implies that

\begin{eqnarray}
qp^{\prime\prime}-pq^{\prime\prime}+c (pp^\prime+qq^{\prime})=0 \\
d(qp^{\prime}-pq^{\prime})+\frac{c}{2} d(p^2+q^2)=0 \\
\frac{qp^{\prime}-pq^{\prime}}{p^2+q^2}+\frac{c}{2} =0 \\
d(\arctan\left[\frac{p}{q}\right])+\frac{c}{2}=0
\end{eqnarray}

From where we determine the solution

\begin{equation*}
\frac{p}{q}=\tan \left( c_{3}-\frac{c}{2}r\right) .
\end{equation*}%
However, for specific values of the free parameters, $c_{1}$, $c_{2},$ $n$
and $m$, the dynamical system (\ref{2.9}), (\ref{2.10}) admits additional
Lie point symmetries.

\subsubsection{Case I: $c_{1}=0$\strut \hfill }

If $c_{1}=0$ then (\ref{2.9}) and (\ref{2.10}) is simplified as%
\begin{eqnarray}
p^{\prime \prime }+cq^{\prime } &=&p\left( \frac{(p^{2}+q^{2})^{m}}{c^{2}-1}%
+c_{2}+(p^{2}+q^{2})^{n}\right)   \label{2.35} \\
q^{\prime \prime }-cp^{\prime } &=&q\left( \frac{(p^{2}+q^{2})^{m}}{c^{2}-1}%
+c_{2}+(p^{2}+q^{2})^{n}\right)   \label{2.36}
\end{eqnarray}

The solutions of (\ref{2.35}) and (\ref{2.36}) are expressed by the functions

\begin{eqnarray}
p&=&R(r) \sin[\theta(r)]  \label{2.33} \\
q&=&R(r) \cos[\theta(r)]  \label{2.34}
\end{eqnarray}

where
\begin{equation*}
\theta (r)=\int \dfrac{c_{3}}{{R(t)}^{2}}dt+\frac{ct}{2}+c_{4}
\end{equation*}%
and $R(t)$ can be found from the following equation

\begin{equation}\label{2.37}
R^{\prime \prime }=-\frac{1}{4}c^{2}R+c_{2}R+\frac{R^{2m+1}}{c^{2}-1}%
+R^{2n+1}+\frac{c_{3}}{R^{3}}
\end{equation}%
where $^{\prime }$ represents the derivative with respect to $r.$

The trivial symmetry of (\ref{2.37}) is $\partial _{r}.$ Therefore the above
equation reduced as follows. Let $R(r)=R$ and $R^{\prime }=\phi (R)$ then (%
\ref{2.37})

\begin{equation}\label{2.38}
\phi \dfrac{d\phi }{dR}=-\frac{1}{4}c^{2}R+c_{2}R+\frac{R^{2m+1}}{c^{2}-1}%
+R^{2n+1}+\frac{{c_{3}}^{2}}{R^{3}}
\end{equation}%
where $\phi $ is a function of $R.$ The solution of (\ref{2.38}) is given by

\begin{equation}
\phi =\pm \sqrt{\left( c_{2}-\frac{c^{2}}{4}\right) R^{2}+\frac{R^{2m+2}}{%
(c^{2}-1)(m+1)}+\frac{R^{2n+2}}{n+1}-\frac{{c_{3}}^{2}}{R^{2}}+c_{5}}
\end{equation}%
that is%
\begin{equation}
\int \frac{dR}{\sqrt{\left( c_{2}-\frac{c^{2}}{4}\right) R^{2}+\frac{R^{2m+2}%
}{(c^{2}-1)(m+1)}+\frac{R^{2n+2}}{n+1}-\frac{{c_{3}}^{2}}{R^{2}}+c_{5}}}%
=r-r_{0}.
\end{equation}

\subsubsection{\strut Case II:~$c_{1}=c_{2}=0~$and $m=n=-2$\hfill }

If $c_{1}=c_{2}=0$ and $m=n=-2$ then equations (\ref{2.9}) and (\ref{2.10})
reduces to

\begin{eqnarray}
p^{\prime\prime}+c q^\prime&=&\left(\frac{c^2}{c^2-1}\right)\frac{p}{%
(p^2+q^2)^{2}}  \label{2.31} \\
q^{\prime\prime}-c p^\prime&=&\left(\frac{c^2}{c^2-1}\right)\frac{q}{%
(p^2+q^2)^{2}}  \label{2.32}
\end{eqnarray}

The symmetries of the above equations are given by

\begin{equation}
\begin{array}{lll}
\Gamma _{7} & = & \partial _{r} \\
\Gamma _{8} & = & p\partial _{q}-q\partial _{p} \\
\Gamma _{9} & = & (cp\sin [cr]+cq\cos [cr])\partial _{p}+(cq\sin [cr]-cp\cos
[cr])\partial _{q}-2\cos [cr]\partial _{r} \\
\Gamma _{10} & = & (cq\sin [cr]-cp\cos [cr])\partial _{p}+(cp\sin
[cr]+cq\cos [cr])\partial _{q}+2\sin [cr]\partial _{r}%
\end{array}%
\end{equation}

Surprisingly, the Lie point symmetries~$\Gamma _{7},~\Gamma _{9}$ and $\Gamma
_{10}$ for the $SL\left( 2,R\right) $ Lie algebra, which means the resulting
system can be reduced to the Ermakov-Pinney system \cite{erm1,erm2,erm3}.

Indeed under, the change of variables
\begin{eqnarray}
p &=&R(r)\sin [\int \dfrac{c_{3}}{{R(t)}^{2}}dt+\frac{ct}{2}+c_{4}]
\label{2.33} \\
q &=&R(r)\cos [\int \dfrac{c_{3}}{{R(t)}^{2}}dt+\frac{ct}{2}+c_{4}]
\label{2.34}
\end{eqnarray}%
we find the analytic solution which is expressed in closed form functions%
\begin{equation*}
R(t)=\pm \dfrac{\sqrt{c_{4}\left( (c^{2}-1)c^{2}+{c_{3}}^{2}c^{2}-{c_{4}}%
^{2}\exp [-2ict]+2cc_{5}{c_{4}}^{2}\exp [-ict]-c^{2}{c_{4}}^{2}{c_{5}}%
^{2}\right) }}{cc_{4}\exp [-\frac{ic}{2}t]}.
\end{equation*}

\subsubsection{\strut Case III: $m=n=-2$\hfill }

If $m=n=-2$ then (\ref{2.9}) and (\ref{2.10}) reduces to

\begin{eqnarray}
p^{\prime\prime}+c q^\prime&=&\left(\frac{c^2}{c^2-1}\right)\frac{p}{%
(p^2+q^2)^{2}} +(c_1r+c_2)p  \label{2.40} \\
q^{\prime\prime}-c p^\prime&=&\left(\frac{c^2}{c^2-1}\right)\frac{q}{%
(p^2+q^2)^{2}} +(c_1r+c_2)q  \label{2.41}
\end{eqnarray}

The solution of (\ref{2.40}) and $(\ref{2.41})$are given by

\begin{eqnarray}
p&=&R(r) \sin[\theta(r)]  \label{2.33} \\
q&=&R(r) \cos[\theta(r)]  \label{2.34}
\end{eqnarray}

where
\begin{equation*}
\theta (r)=\int \dfrac{c_{3}}{{R(t)}^{2}}dt+\frac{ct}{2}+c_{4}
\end{equation*}
and $R(t)$ can be found from the following equation

\begin{equation*}
R^{\prime \prime }=(c_{1}r+c_{2}-\frac{c^{2}}{4})R+\left( \frac{c^{2}}{%
(c^{2}-1)}+{c_{3}}^{2}\right) \left( \frac{1}{R^{3}}\right)
\end{equation*}%
the latter equation is nothing else than the Ermakov-Pinney equation, while
its symmetries are

\begin{equation}
\begin{array}{lll}
\Gamma _{11} & = & {c_{1}}^{\frac{1}{3}}AiryAi\left[ \frac{%
4(c_{1}r+c_{2})-c^{2}}{4{c_{1}}^{\frac{2}{3}}}\right] AiryAi^{\prime }\left[
\frac{4(c_{1}r+c_{2})-c^{2}}{4{c_{1}}^{\frac{2}{3}}}\right] R\partial _{R}%
\nonumber \\
&  & +AiryAi\left[ \frac{4(c_{1}r+c_{2})-c^{2}}{4{c_{1}}^{\frac{2}{3}}}%
\right] ^{2}\partial _{r} \\
\Gamma _{12} & = & {c_{1}}^{\frac{1}{3}}AiryBi\left[ \frac{%
4(c_{1}r+c_{2})-c^{2}}{4{c_{1}}^{\frac{2}{3}}}\right] AiryBi^{\prime }\left[
\frac{4(c_{1}r+c_{2})-c^{2}}{4{c_{1}}^{\frac{2}{3}}}\right] R\partial _{R}%
\nonumber \\
&  & +AiryBi\left[ \frac{4(c_{1}r+c_{2})-c^{2}}{4{c_{1}}^{\frac{2}{3}}}%
\right] ^{2}\partial _{r} \\
\Gamma _{13} & = & {c_{1}}^{\frac{1}{3}}\left( AiryAi\left[ \frac{%
4(c_{1}r+c_{2})-c^{2}}{4{c_{1}}^{\frac{2}{3}}}\right] AiryBi^{\prime }\left[
\frac{4(c_{1}r+c_{2})-c^{2}}{4{c_{1}}^{\frac{2}{3}}}\right] \right.  \\
&  & +\left. AiryBi\left[ \frac{4(c_{1}r+c_{2})-c^{2}}{4{c_{1}}^{\frac{2}{3}}%
}\right] AiryAi^{\prime }\left[ \frac{4(c_{1}r+c_{2})-c^{2}}{4{c_{1}}^{\frac{%
2}{3}}}\right] \right) R\partial _{R}\nonumber \\
&  & +2AiryAi\left[ \frac{4(c_{1}r+c_{2})-c^{2}}{4{c_{1}}^{\frac{2}{3}}}%
\right] AiryBi\left[ \frac{4(c_{1}r+c_{2})-c^{2}}{4{c_{1}}^{\frac{2}{3}}}%
\right] \partial _{r}%
\end{array}%
\end{equation}

Finally, the solution of (\ref{2.41}) is expressed in terms of the Airy
function as follows

\begin{equation}
R(r)=c_6 AiryAi\left[\frac{4(c_1r+c_2) -c^2}{4{c_1}^{\frac{2}{3}}}\right]
\end{equation}

with a condition $c_{3}=\pm \dfrac{c}{\sqrt{1-c^{2}}}.$

\strut\hfill

\section {Conclusion}

The algebraic properties of the generalized Zakharov equations was the subject of this study. We wrote the Zakharov equations as a system of three differential equations of real values functions.
For the latter system we determined the infinitesimal transformations which leave invariants the Zakharov equations. The admitted Lie point symmetries are five. From the Lie symmetries we found a similarity transformation which provides reductions which lead to travel-wave solutions.
The reduced system has been classified according to the admitted Lie point symmetries, where we found surprisingly that the Ermakov-Pinney equation can describe the generalized Zakharov equations.
This work demonstrate the application of the Lie point symmetries in applied mathematics and in example in plasma physics. In a future work we plan to investigate the nature of the conservation laws provided by the symmetry vectors.

\section*{Acknowledgements}
 KK thank Prof. Dr. Stylianos Dimas, S\'ao Jos\'e dos Campos/SP, Brasil for providing us new version of SYM-Package.

\end{document}